\documentclass[sn-mathphys, iicol, referee]{sn-jnl}% Math
\jyear{2021}%

\theoremstyle{thmstyleone}%
\theoremstyle{thmstyletwo}%

\theoremstyle{thmstylethree}%

\usepackage{graphicx}
\usepackage{natbib}
\usepackage{dcolumn}   % needed for some tables
\usepackage{bm}        % for math
\usepackage{verbatim}   % for math
\usepackage{textcomp}
\usepackage{amssymb}
\usepackage{amsmath}
\usepackage{url}
\usepackage{natbib}
\usepackage{multicol,lipsum}
\usepackage{cuted}  %%write long eqs
\usepackage{flushend}

\begin{document}

\title[Article Title]{Simulation of the RIBRAS Facility with GEANT4}

\author[1]{\fnm{L. E.} \sur{Tamayose}}
 \email{leoeiji@usp.br}

\author*[1]{\fnm{J. C.} \sur{Zamora}}\email{zamora@nscl.msu.edu}

\author[1]{\fnm{G. F. } \sur{Fortino}}

\author[2]{\fnm{D.} \sur{Flechas}}

\affil[1]{\orgdiv{Instituto de F\'isica}, \orgname{ Universidade de S\~ao Paulo}, \orgaddress{\city{S\~ao Paulo}, \country{Brazil}}}

\affil[2]{\orgdiv{Departamento de F\'isica}, \orgname{Universidad Nacional de Colombia}, \orgaddress{ \city{Bogot\'a}, \country{Colombia}}}


\abstract{A \textsc{Geant4} simulation code was developed to perform realistic simulations of the RIBRAS facility. A second order expansion of a finite solenoid field was included to describe  the beam optics with a good precision. A systematic study of coil currents for several magnetic rigidities and focal points was performed. Parameterizations of the coil currents for single and dual mode operations were obtained. Dedicated routines were developed to simulate the mechanism of  direct reactions involving two and three particles in the final state. The present simulations were employed to investigate the feasibility of a Solenoidal Spectrometer with the RIBRAS facility. Our first results indicate that the concept can be applied in the RIBRAS system under certain conditions.  Forthcoming studies both from simulations and experiment are already under development.}




\maketitle

\section{Introduction}
Double superconducting solenoid systems for rare-isotope beams production such as RIBRAS \cite{ribras1,ribras2}  and TWINSOL \cite{twinsol1,twinsol2} have been used for more than two decades in studies of nuclear reactions with radioactive beams at energies near  the Coulomb barrier. The strong magnetic fields operated in the solenoids provide a compact and efficient method for secondary beam production and selection based on the magnetic rigidity. \par
 Dedicated magnetic field simulations are required in order to  describe the particle trajectories and to define the best position of blockers and degraders. The \textsc{Geant4} \cite{geant4} toolkit  offers an excellent method to develop very precise simulations by using the diverse functionalities available for particle transport and its interaction with matter. Realistic simulations in 3D are possible with \textsc{Geant4} due its versatile geometry definition and to the methods used to integrate the particle equation of motion. The toolkit allows the implementation of new routines where it is possible to include specific reaction mechanisms at energies near the Coulomb barrier. Future projects and new experimental developments can be investigated with the simulations. One of these projects is the feasibility of a Solenoidal Spectrometer with the RIBRAS facility that is discussed in this paper.

\section{The RIBRAS Facility}
\begin{figure}[!ht]
\centering
\includegraphics[width=0.5\textwidth]{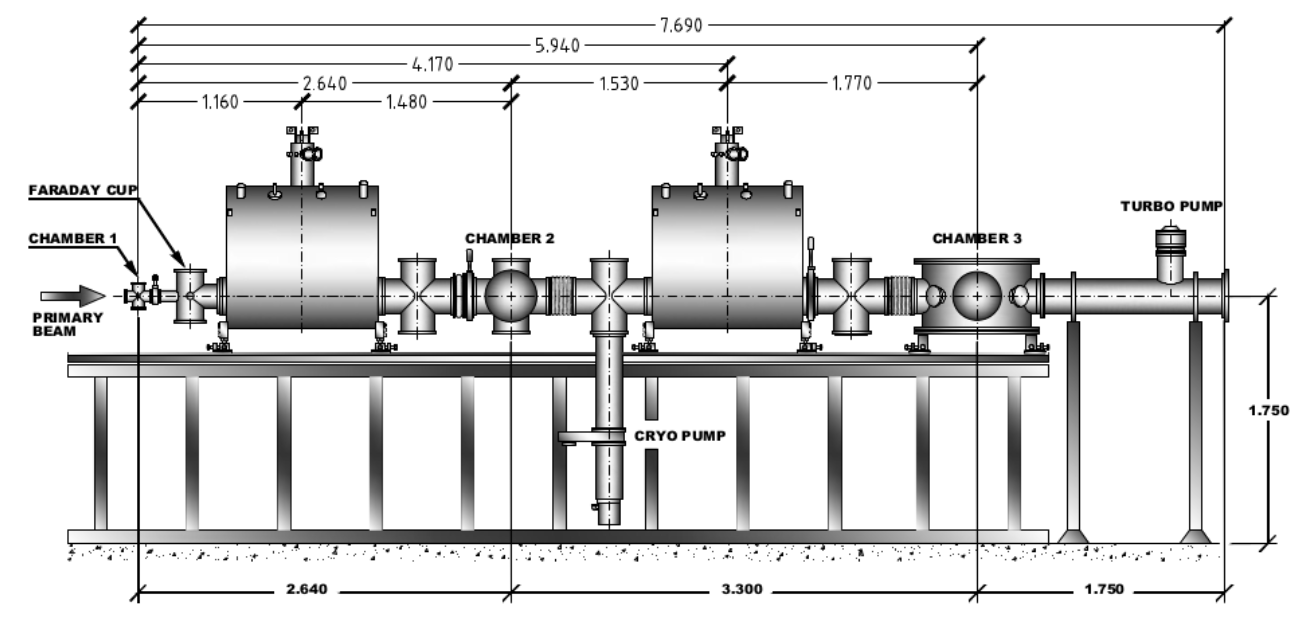}
\caption{\label{ribras} Sketch of the RIBRAS facility. Two superconducting solenoids are used to select and focus radioactive-ion beams in the scattering chambers. In the single-solenoid mode, the particles are focus in the central scattering chamber (chamber 2). A larger scattering chamber (chamber 3) is used for the dual-solenoid mode.  }
\end{figure}

The RIBRAS (Radioactive Ion Beam in Brazil)  facility is a double superconducting solenoid system for rare-isotope beams production  The concept is similar to the TWINSOL facility. Both systems use an in-flight magnetic rigidity ($B\rho$) selection of reaction products emerging from a primary target. A sketch of RIBRAS is shown in Figure.~\ref{ribras}.

The superconducting solenoids have a maximum field of 6.5~T and a 30~cm clear warm bore where the beam-line passes through. Secondary beams are produced in the primary target, which is located in front  of the first magnet (see Fig.~\ref{ribras}). The angular acceptance is in the range from $2^\circ$ to $15^\circ$ in the laboratory system. The lower angular limit is defined by a Faraday cup inserted at $0^\circ$, while the upper limit is given by the beam pipe diameter. The solenoids act as thick lens focusing ions with a given $B\rho$ to a secondary target located downstream in a scattering chamber. In the single-solenoid mode, the focal point corresponds to the center of the middle scattering chamber (chamber 2 in Fig.~\ref{ribras}) at 2.6~m distance from the production target. Thus,  the strength of the magnetic field is adjusted to focus the secondary ion beams with a certain magnetic rigidity at the reaction target that is located in the focal point.  The double-solenoid mode provides a  better selectivity of the beam particles by using also a degrader and a set of blockers  along the flying path. The focal point in the double-solenoid mode is at 5.9~m away from the primary target, located in the center of the scattering chamber 3 (see Fig.~\ref{ribras}). \par
In order to appropriately simulate the beam optics, the particle equation of motion in a realistic magnetic field has to be calculated numerically. A program developed for the TWINSOL setup \cite{phd_twinsol} has been adapted to simulate the beam optics of the RIBRAS facility. However, the program only provides information of 2D projections of the particle trajectories. In the present work, we have extended this study for three dimensions. Now, our program allows to investigate in a better detail the particle trajectories, asymmetry effects, reactions in the solenoids and many more. In the next Section, it is explained how the beam optics was implemented in the simulation.

\section{Beam Optics}

\subsection{Magnetic Field}
In a simple approximation, the magnetic field of a solenoid  can be assumed to be zero outside of the solenoid region and uniform inside it. When a charged particle enters to an uniform magnetic field, its trajectory is affected by the Lorentz force that makes the particle rotate around the solenoid axis in an helicoid form. The oscillation of this trajectory is defined by the cyclotron frequency \cite{magnet1}
\begin{equation}
 \omega_\text{cyc} = \frac{eqB}{\gamma m},
\end{equation}
where  $eq$ is the particle charge, $B$ is the magnetic-field strength, $\gamma$ is the relativistic Lorentz factor and $m$ is the particle mass. However, this simple model is not able to describe the focusing effect of a solenoid. A more realistic model assuming the finite size of the solenoid is applied to calculate the off-axis field that focus the beam particles. In general, the components of the axial symmetric magnetic field of a solenoid can be expanded in a Taylor series as \cite{magnet1}
\begin{equation}
\label{eq1}
 B_z(z,r) = B(z) - \frac{r^2}{2}B''(z)+\cdots,
 \end{equation}

\begin{equation}
\label{eq2}
 B_r(z,r) = -\frac{r}{2}B'(z) + \frac{r^3}{16}B'''(z)+\cdots,
 \end{equation}
where $z$ is the position along the solenoid axis, $r$ is the radial distance from the solenoid axis and the prime denotes the derivative with respect to $z$. The radial part provides an impulse to the particle at the solenoid edges that curves its trajectory, while the longitudinal component keeps the particle in a helical trajectory. In our simulation, we used an expansion  up to second order to achieve a good precision in the particle trajectories. Thus,  Eqs.~(\ref{eq1}) and (\ref{eq2}) can be expressed by \cite{phd_twinsol,magnet2, magnet3}

\begin{figure}[!ht]
\centering
\includegraphics[width=0.5\textwidth]{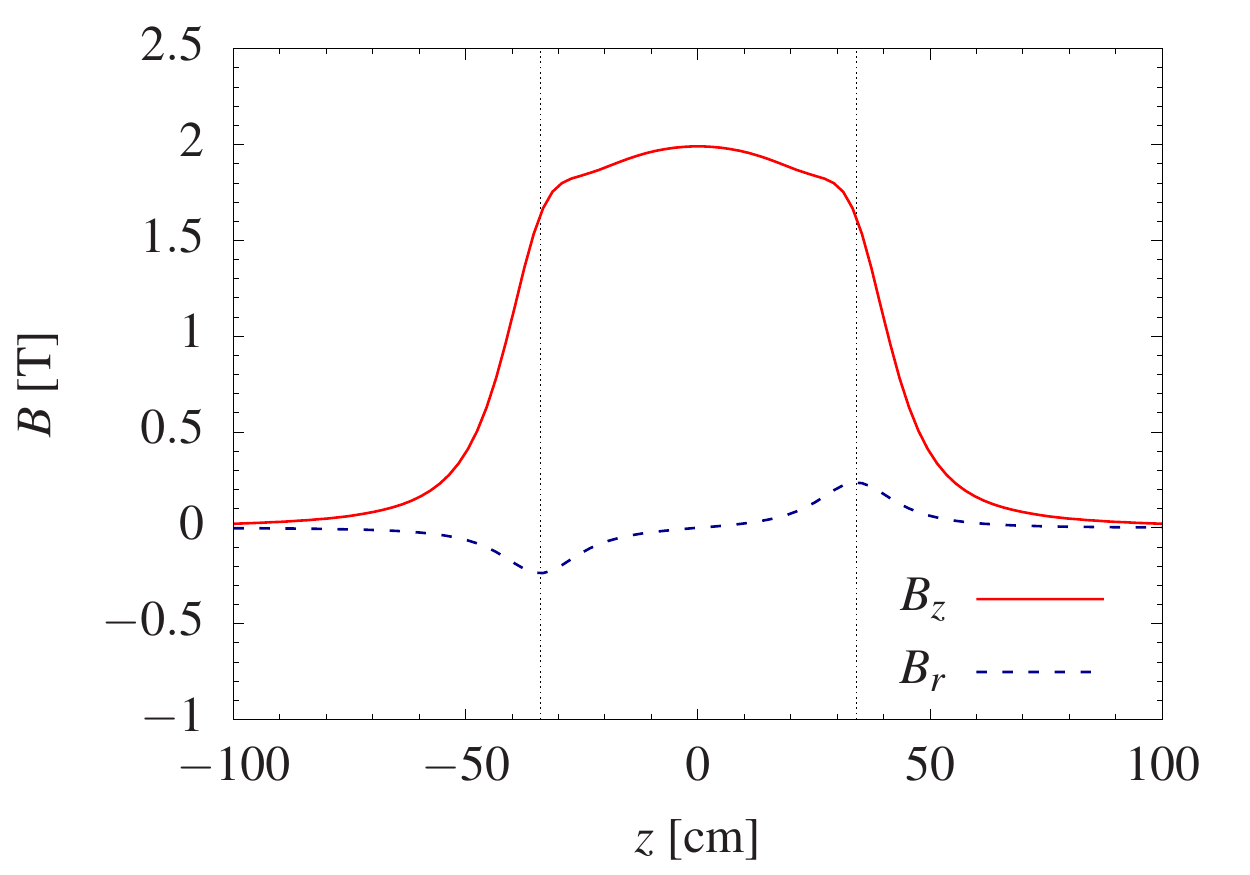}
\caption{\label{bfield} (color online). Off-axis magnetic field components in the longitudinal coordinate ($z$)  for a coil current of 28 Ampere. The vertical dash-dotted lines represent the coil edges.  }
\end{figure}

   \begin{strip}
\begin{equation}
\label{eq3}
 B_z(z,r) = B_0 \frac{\sqrt{R^2+\frac{1}{4}L^2}}{L}  \left[  \left( \frac{z_-}{(R^2 + z_-^2)^{1/2}} -\frac{z_+}{(R^2 + z_+^2)^{1/2}}    \right) - \frac{3}{4}r^2R^2 \left( \frac{z_+}{(R^2 + z_-^2)^{5/2}} -\frac{z_-}{(R^2 + z_+^2)^{5/2}}    \right)  \right],
 \end{equation}

 \begin{equation}
 \label{eq4}
 B_r(z,r) = B_0 \frac{R^2 \sqrt{R^2+\frac{1}{4}L^2}}{L}  \left[  -\frac{r}{2} \left( \frac{1}{(R^2 + z_-^2)^{3/2}} -\frac{1}{(R^2 + z_+^2)^{3/2}}    \right) + \frac{3}{8}r^3 \left( \frac{4z_-^2-R^2}{(R^2 + z_-^2)^{3/2}} -\frac{4z_+^2-R^2}{(R^2 + z_+^2)^{3/2}}    \right)  \right],
 \end{equation}
\end{strip}

where $L$ is the solenoid length, $R$ is the bore radius and $z_\pm = z \pm \frac{L}{2}$. The parameter $B_0=kI$ corresponds to the axial field at the center of the magnet, which is proportional to the electric current ($I$) in the solenoid coil and to the intrinsic constant ($k$) of the  solenoid magnet. The latter value is extracted from the experimental field map measured for each solenoid. Figure~\ref{bfield} shows an example of the magnetic field components as a function of the longitudinal coordinate for $r>0$. The $B_z$ component is nearly constant inside the solenoid coil (vertical dash-dotted lines), and strongly decays in the outside region. The intensity of the radial component is maximum at the solenoid edges, but with opposite sign due to  the magnetic field poles. In the inner region of the solenoid, the radial part is negligible and the field is dominated by the longitudinal component.\par
A non-uniform magnetic field was created in the simulation by using the \textsc{G4MagneticField} class with the parameterizations of Eqs.~(\ref{eq3}) and (\ref{eq4}) to define a field inside and outside of the solenoid volume. This class can easily be replicated to more than one solenoid, as explained below.

\subsection{Single-Solenoid Mode}
The single-solenoid mode corresponds to a setup with the first solenoid that is in front of the production target. Thus,  secondary particles are propagated in the non-uniform field that was previously defined. The trajectories are obtained by solving the respective equation of motion of the particle in the magnetic field. \textsc{Geant4} uses a  Runge-Kutta algorithm to integrate numerically the ordinary differential equations of motion. In our simulation, a fourth order Runge-Kutta method  was used with an accuracy  of 1~$\mu$m for the integrated steps. \par

\begin{figure}[!ht]
\centering
\includegraphics[width=0.35\textwidth]{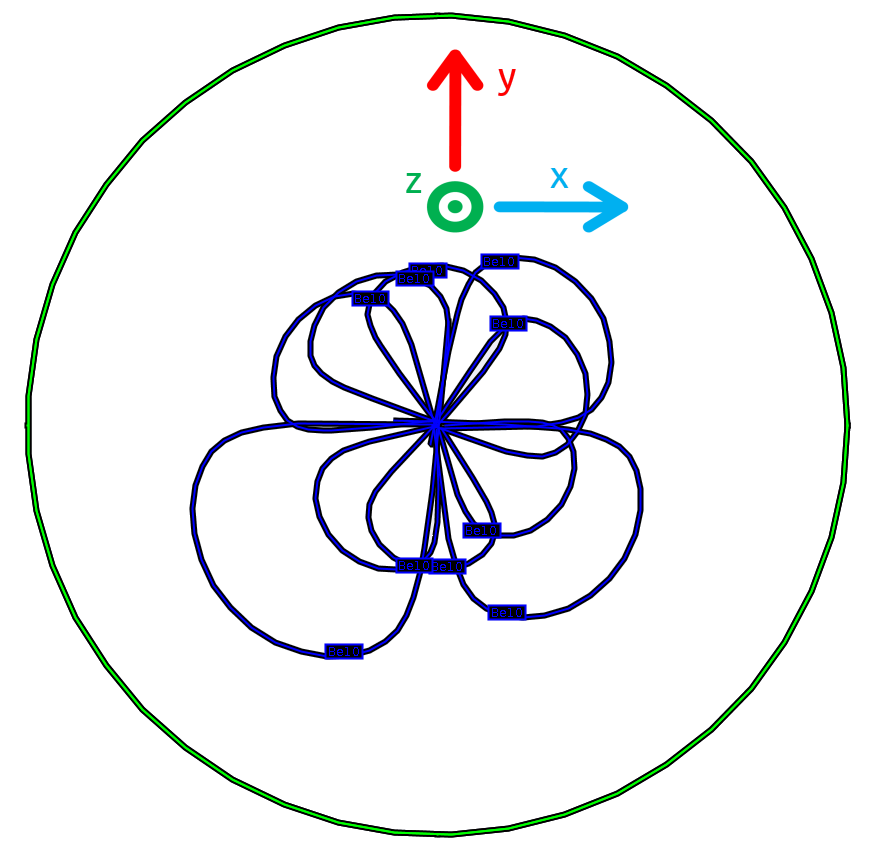}
\caption{\label{rfield} (color online). Helical trajectories (2D projection) observed from the production target. The outer circle is the solenoid bore. Each particle trajectory has an inserted label. }
\end{figure}

The particles that are propagated in the magnetic field  have helical trajectories that cross the solenoid axis, as shown in Figure~\ref{rfield}. The outer circle is the solenoid bore and each blue line correspond to the trajectory of different particles around the $z$ axis. As can be seen, all the trajectories have the same initial and final positions due to the beam focus.

\begin{figure}[!ht]
\centering
\includegraphics[width=0.45\textwidth]{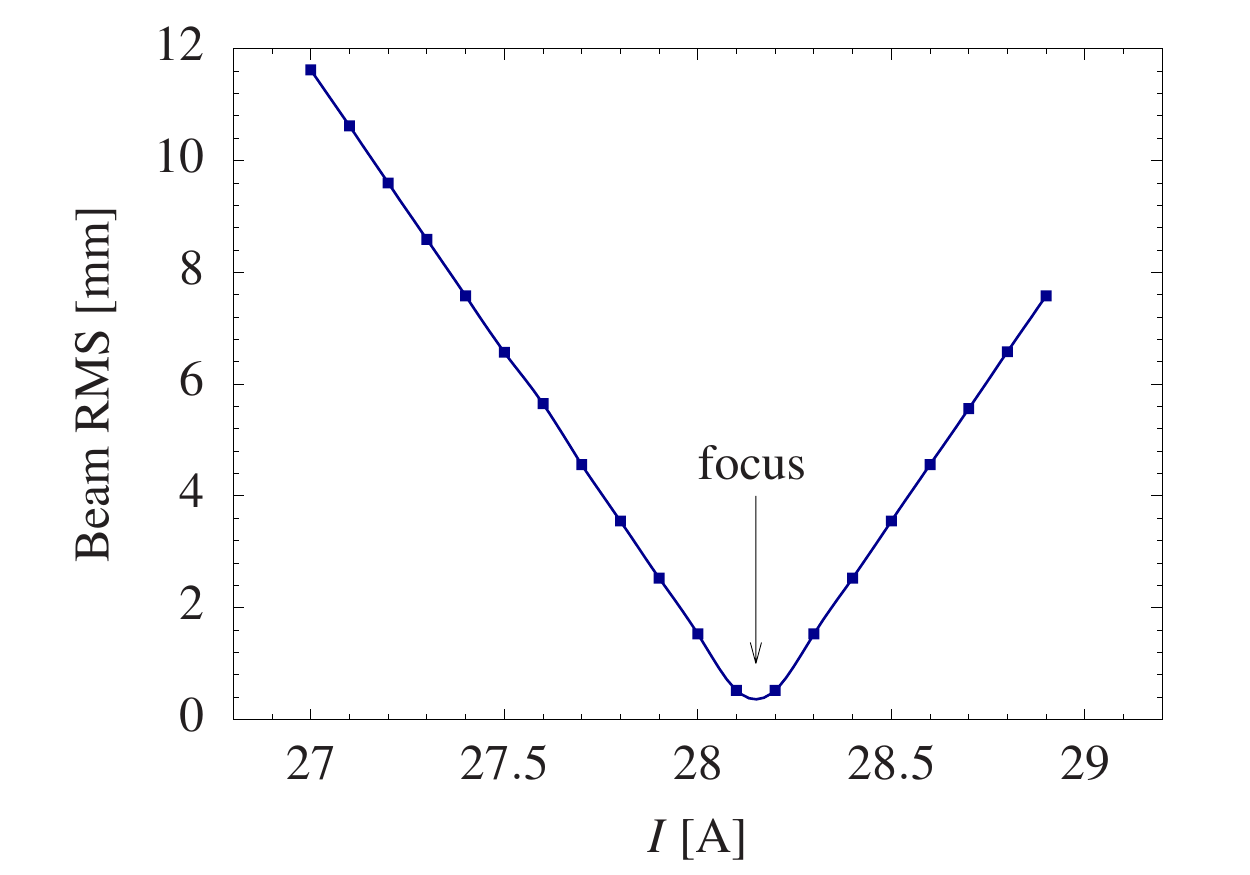}
\includegraphics[width=0.45\textwidth]{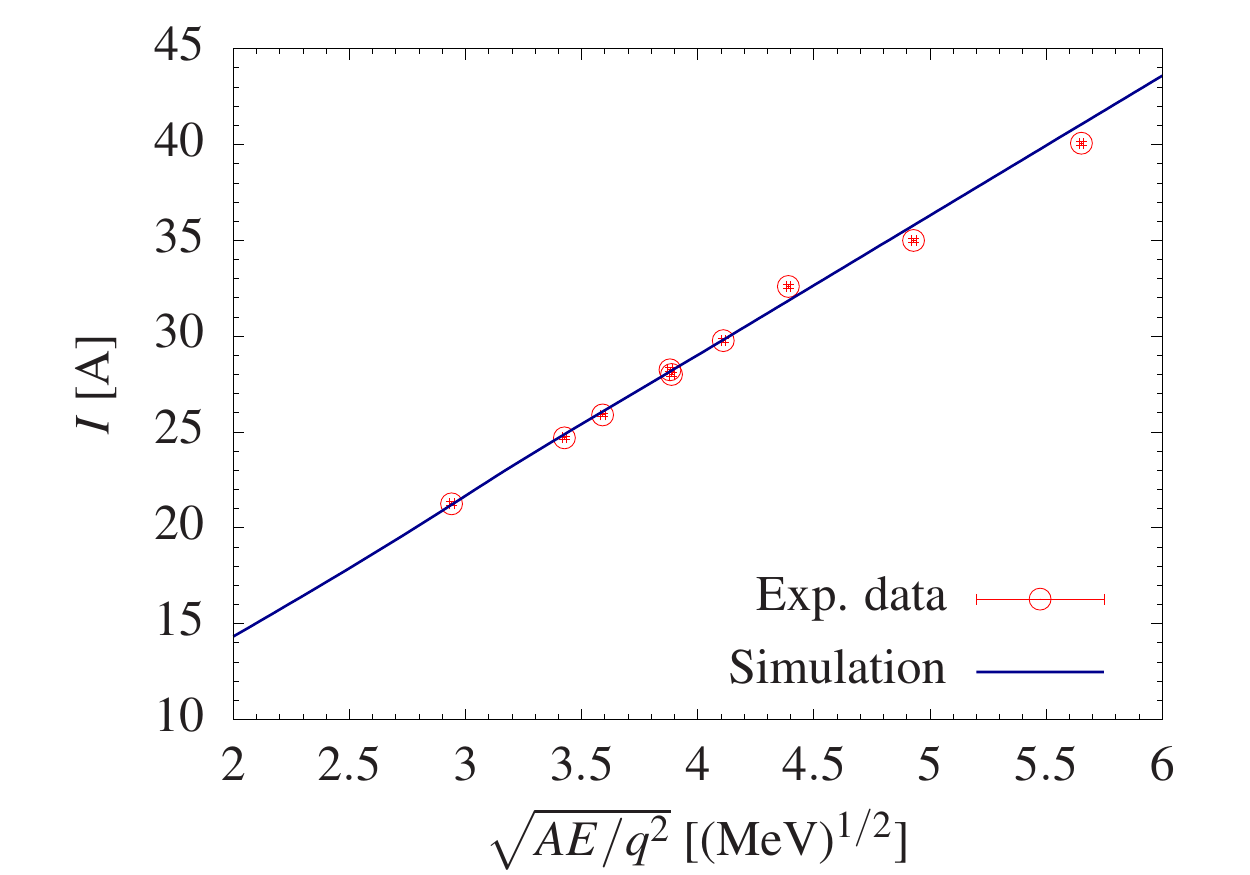}
\caption{\label{focus1s} (color online). Beam focusing for the single-solenoid mode. (Top) Beam RMS radius at the target position for several electric currents in the coil. (Bottom) Coil current for the beam focus as a function of the magnetic rigidity. }
\end{figure}

In the present case, the focal point is located at the center of the scattering chamber 2 (see Fig.~\ref{ribras}). In an experiment, the beam focusing is performed by sweeping the electric current of the magnet to find the maximum  beam intensity at the reaction target. Similarly, hundreds of simulations were performed for different coil currents to obtain the best value that minimize the beam radius in the focal plane. Figure~\ref{focus1s} (top) shows an example of a beam focus using several simulations with a step of 0.1 Ampere. The focus corresponds to the minimum beam RMS (root mean square) radius at the reaction target. The procedure was repeated for many beams covering a wide range of magnetic rigidities. As can be observed in Figure~\ref{focus1s} (bottom), the simulated result has a linear trend in the studied magnetic rigidity region. The simulation is in very good agreement with the solenoid currents obtained in several experiments with different beams (Ref.~\cite{zamora_mater} and references therein). The parameterization for the simulated current (in Ampere) is
\begin{equation}
 I = 7.28\sqrt{\frac{AE}{q^2}}-0.09,
\end{equation}
where $A$ is the mass number of the beam particle, $E$ is the kinetic energy (in MeV) and $q$ is the charge-state number. This linear approximation allows to extract in a very simple way the best solenoid current without having to run a new simulation.

\subsection{Dual-Solenoid Mode}
\begin{figure}[!ht]
\centering
\includegraphics[width=0.5\textwidth]{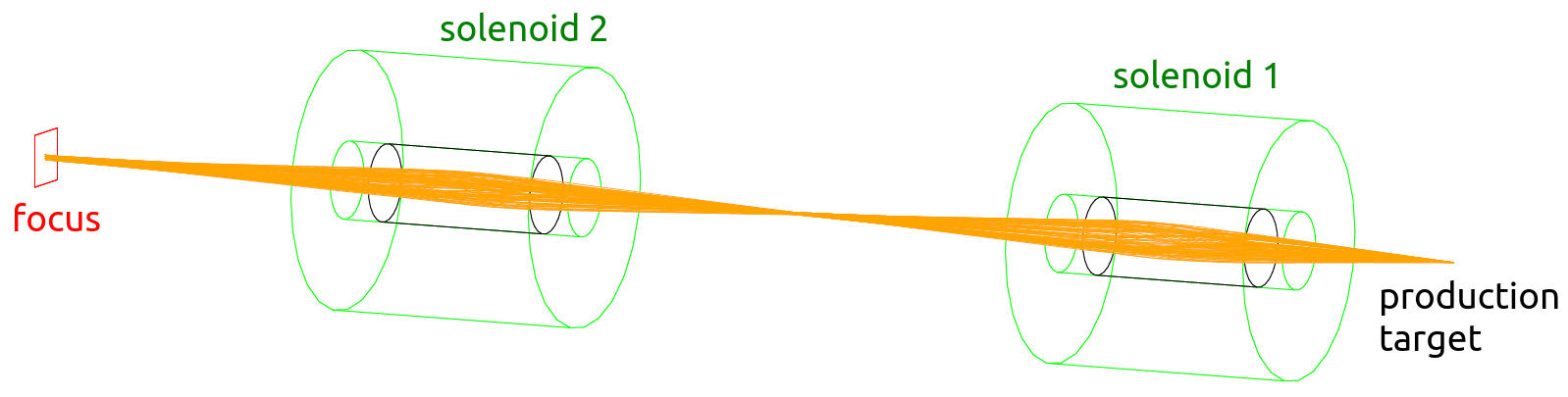}\\
\includegraphics[width=0.5\textwidth]{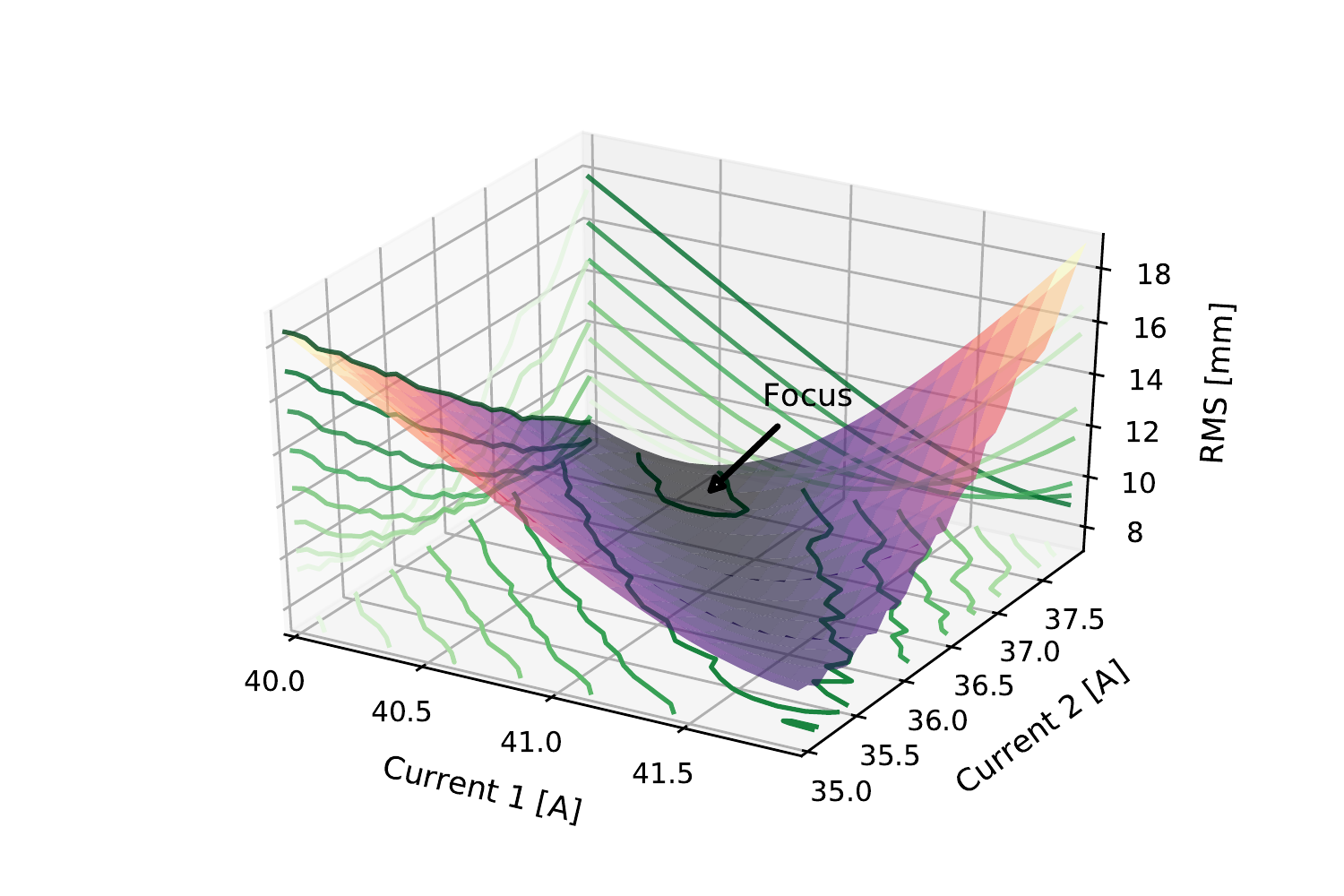}
\caption{\label{focus2s} (color online). Beam focusing for the dual-solenoid mode. (Top) Example of the simulated trajectories with a beam focus. (Bottom) A 3D map of the solenoid currents to obtain a global minimum that corresponds to the beam focus. }
\end{figure}

The \textsc{G4MagneticField} class was extended to use a non-uniform field in two solenoids. The magnetic field definition allows to use solenoids with different coil currents and constant factors. The magnetic field is also defined in the outer region of the solenoids. In this configuration, a  superposition of the magnetic fields is included to integrate the particle trajectory across the two solenoids setup. The secondary particles emerging from the production  target are focus by the first solenoid in the chamber 2, where it is possible to include a beam degrader. The particles that are accepted in the second solenoid are focus in the scattering chamber 3 that is at 5.9~m away from the production target (Fig.~\ref{ribras}). Figure~\ref{focus2s} (top) shows an example of the simulated particle trajectories with the dual-solenoid mode. Given that the two solenoids have independent strengths, the phase-space covered to parameterize the two coil currents (for any $B\rho$ value) is considerably larger than for the single-solenoid mode. Thousands of simulations varying the solenoid currents, $I_1$ and $I_2$, were performed  to find the best configuration that minimize the beam radius at the reaction target. Figure~\ref{focus2s} (bottom) shows an example of the phase-space covered to find the beam focus for a given $B\rho$ value.

Similar to the single-solenoid mode, the minimization procedure was performed for several magnetic rigidities in order to obtain a parameterization. Also, the simulation result for the two solenoids exhibit a linear trend that are given by
\begin{equation}
 I_1 = 6.59\sqrt{\frac{AE}{q^2}}+2.88,
\end{equation}

\begin{equation}
 I_2 = 5.50\sqrt{\frac{AE}{q^2}}+2.25,
\end{equation}
where the subindex indicates the solenoid number. The parameterization of above provides an educated guess for the focal currents, which can be used during an experiment to find the best coil currents in a more efficient way.

\section{Event Generator}
\begin{figure}[!ht]
\centering
\includegraphics[width=0.5\textwidth]{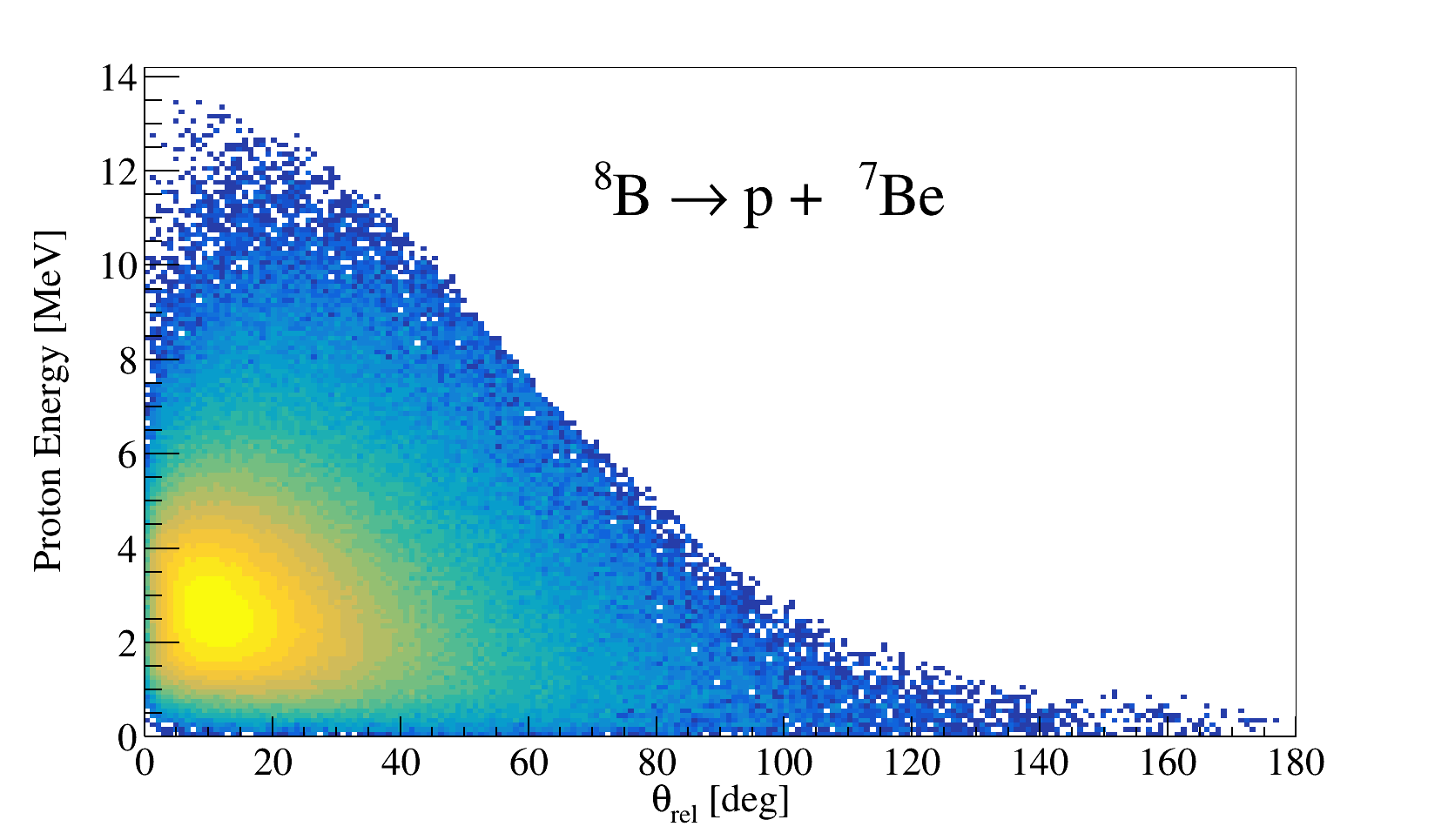}
\caption{\label{kin3b} (color online). Reaction kinematics of ${}^{8}$B breakup on a ${}^{58}$Ni target at 26~MeV. Proton energy vs. the relative angle between  proton and ${}^{7}$Be particles. }
\end{figure}

A nuclear reaction class was developed in the simulation to force a certain reaction process when the particle interacts with a target. The reaction class is based on the \textsc{ G4VParticleChange} method that uses the whole physical information of a primary particle to generate a secondary one using a given model. Relativistic two and three body kinematics generators were implemented using the Mandelstam variables \cite{byckling1973}. In this representation, scattering angles, energy and momentum of the particles can be derived in a compact and complete way in a Lorentz-invariant fashion. Several reaction processes are included in the class, such as elastic/inelastic scattering, pickup/stripping reactions, knockout, breakup, in-flight decay, etc. The  reaction mechanisms can be easily defined in an input macro. Additionally, cross section tables can be included  to generate the reaction events based theoretical angular and energy distributions. As an example, Figure~\ref{kin3b} presents a three body output channel with  ${}^{8}$B breakup   on a ${}^{58}$Ni target at 26~MeV. The kinematics plot shows in the x-axis the relative angle ($\theta_\text{rel}$) between ${}^{7}$Be and proton, and in the y-axis the proton kinetic energy. Cross section tables are included in order to have a more realistic reaction generator.

\section{Solenoidal Spectrometer}
One of the main objectives of the implementation of a \textsc{Geant4} simulation of the RIBRAS facility was to study the feasibility of the system to operate as a Solenoidal Spectrometer. The concept was demonstrated by the HELical Orbit Spectrometer (HELIOS) at Argonne National Laboratory \cite{WUOSMAA20071290,LIGHTHALL201097}. The goal of this experiment is to investigate nuclear reactions in inverse kinematics using a target and detector setup inside of an uniform magnetic field. Several projects around the world with the same concept are under development, such as SOLARIS \cite{solaris}, ISS \cite{iss} and SSNAP \cite{ALLEN2020161350}. \par

Reaction target and detectors are mounted inside of a solenoid. The beam particles impinge the target and induce nuclear reactions. Thus,  recoiling particles produced in the reactions are transported in helical orbits back to the axis where they are detected in a position-sensitive silicon array. After one cyclotron period, $T_\text{cyc} = 2\pi /\omega_\text{cyc}$, the particle returns to the axis at a distance $z$ from the target. Using this information, it is possible to correlate the detected position $z$ with the deposited energy  by \cite{WUOSMAA20071290}
\begin{equation}
\label{sseq}
E_\text{lab} = E_\text{c.m.} - \frac{1}{2}m V^2_\text{c.m.} + \left( \frac{m V_\text{c.m.}}{T_\text{cyc}} \right)z.
\end{equation}
$E_\text{lab}$ is the detected energy (laboratory system), $E_\text{c.m.}$ is the particle energy in the center-of-mass frame and $V_\text{c.m.}$ is the center-of-mass velocity that is fixed by the beam energy. Therefore, different excitation are represented by linear bands in a $E_\text{lab}$-$z$ spectrum. \par

\begin{figure}[!ht]
\centering
\includegraphics[width=0.45\textwidth]{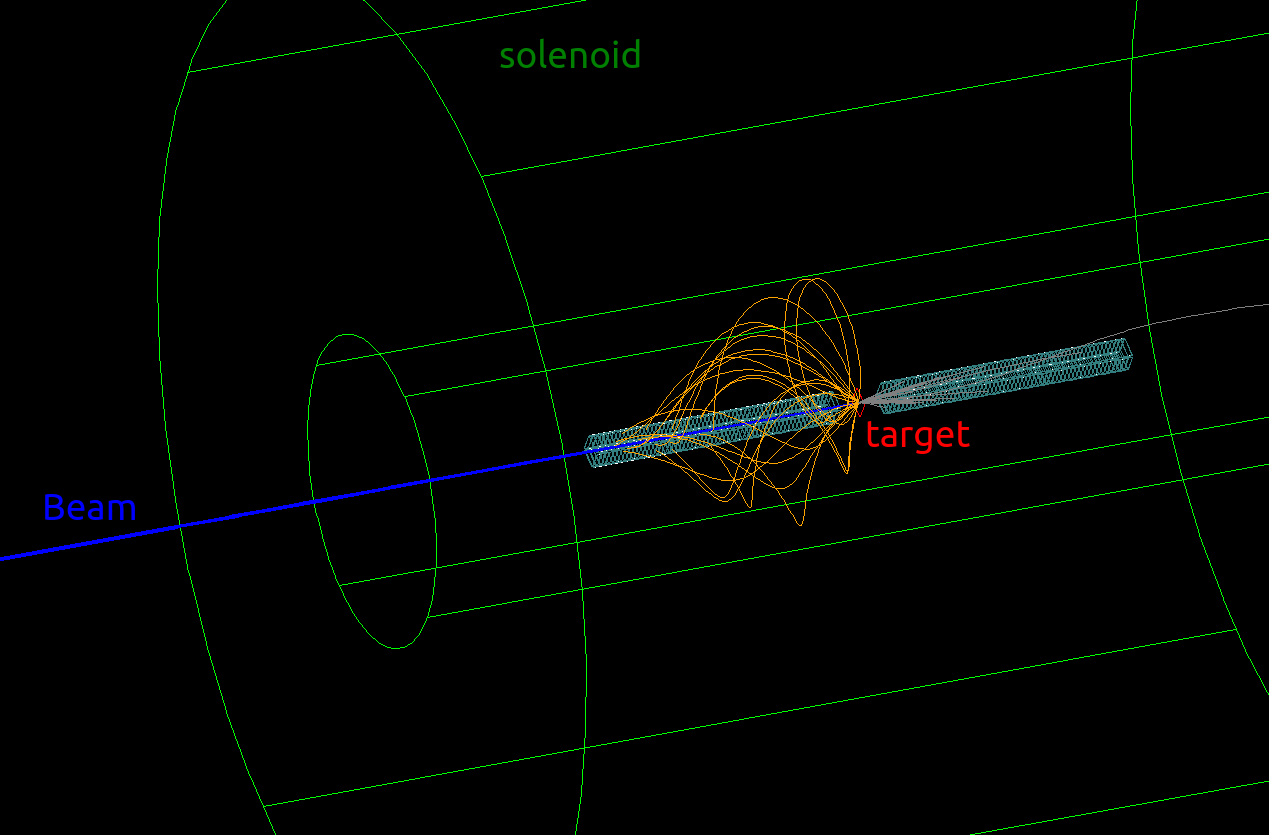}\\
\includegraphics[width=0.5\textwidth]{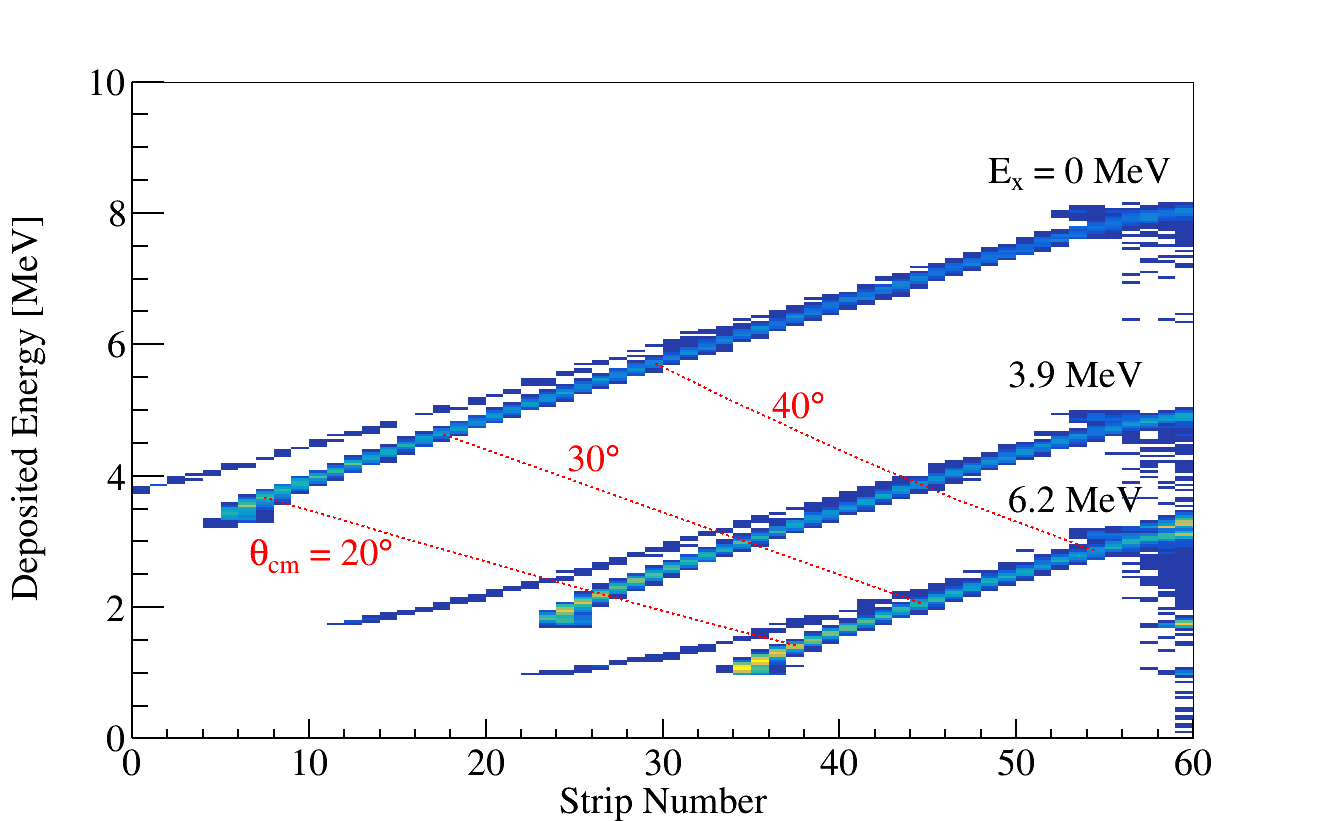}
\caption{\label{ss1} (color online). Simulation of a Solenoidal Spectrometer with only one solenoid of RIBRAS. (Top) Simulation of the recoiling particle trajectories for the reaction ${}^{6}\text{Li}({}^{12}\text{C},\alpha){}^{14}\text{N}$ at 20~MeV. The particles hit the detectors at different positions depending on the center-of-mass angle in which they were produced. (Bottom) Spectrum of the particle deposited energy vs. detector strip number (proportional to the z position). The ground state and the two first $1^+$ states are clearly separated by energy and center-of-mass angles.  }
\end{figure}

The simulated setup of the RIBRAS Solenoid Spectrometer is comprised by 8 position-sensitive silicon detectors of $2\times25$~cm$^2$ area with 60 independent strips. The setup is divided in two groups of 4 detectors symmetrically mounted around the reaction target inside the solenoid, as shown in Figure~\ref{ss1} (top). The detector positions are adjusted to allow the free path of the beam particles, and at the same time to cover a large solid angle. In a single-solenoid mode, the beam particles pass through the hollow of the silicon array and intercept the reaction target. This configuration is adequate for studies of nuclear reactions with stable beams. An example of reaction that can be studied with this setup is the $pn$ transfer reaction in inverse kinematics with a  ${}^{12}\text{C}$ beam on a  ${}^{6}\text{Li}$ target. The recoiling $\alpha$ particles are emitted at backward angles due to the positive $Q$ value. As observed in Figure~\ref{ss1} (top), the recoils are transported in helical orbits to different positions of the detector. The strength of the magnetic field in this case was 6~T. Transfer reactions populating the ${}^{14}\text{N}$ ground state and the two first $1^+$ states (3.9 MeV and 6.2 MeV) were simulated. Figure~\ref{ss1} (bottom) shows the reconstructed spectrum of deposited energy vs. strip number (proportional to the $z$ position). As expected from Eq.~(\ref{sseq}), different linear bands are obtained for the excited states. Also, the $z$ position can be converted to  center-of-mass angle ($\theta_\text{c.m.})$. Even though  a few trajectories are not detected in the right position. This effect is visible in the thin lines close to each band. Possibly geometric effect are producing a separation in the kinematic bands. However, it can be solved by increasing the detector size and distance from the target.

\begin{figure}[!ht]
\centering
\includegraphics[width=0.45\textwidth]{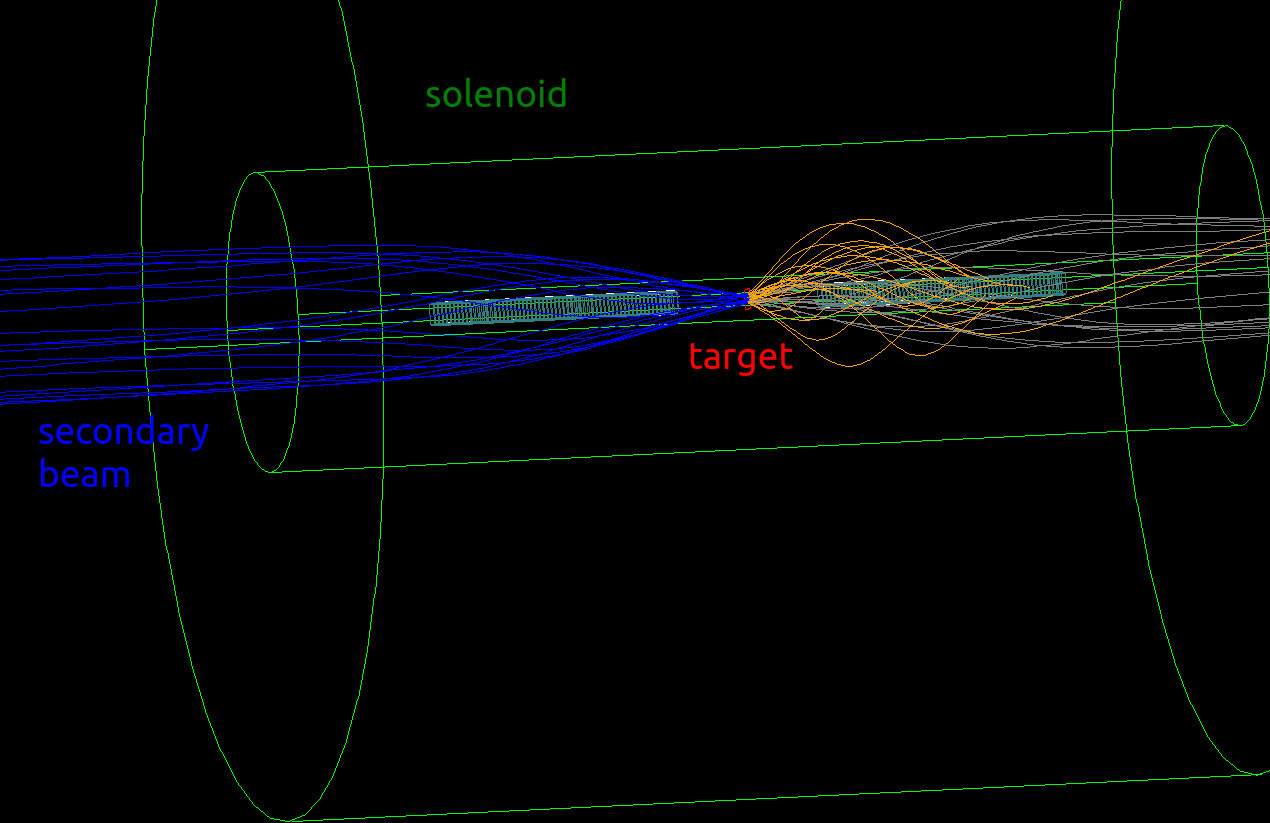}
\caption{\label{ss2} (color online). Simulation of a Solenoidal Spectrometer with the two solenoids of RIBRAS. The dual-solenoid mode is used to investigate nuclear reactions with secondary radioactive beams. The first solenoid focus the beam in the center of the second solenoid where the detectors and target are mounted. The simulated reaction was $p({}^{10}\text{Be},d){}^{9}\text{Be}$. Deuterons are produced in forward direction and detected in the downstream detector setup. }
\end{figure}

In the dual-solenoid mode, the detector array is  placed into the second solenoid,  using the first to focus the beam on the reaction target. This configuration is important for studies with radioactive beams. Figure~\ref{ss2} shows and example of simulated particle trajectories with a ${}^{10}$Be secondary beam. As can be noted, the beam enters to the second solenoid almost parallel, but due to the strong magnetic field the particles are focus on the target. In this case, the reaction studied was $(p,d)$ which have a negative Q value. The recoiling deuterons are transported in helical orbits to the downstream detectors.  Eq.~(\ref{sseq}) is also valid for this example. Further studies are needed to optimize the position of the target and detectors for the dual-solenoid mode with other beams.

\section{Summary}
A \textsc{Geant4} simulation code was developed to simulate the RIBRAS facility. A dedicated class for the magnetic field of the superconducting solenoids was implemented to describe with a good precision the beam optics. The simulation allows to study the particle trajectories in three dimensions, which is of great importance to investigate optical effects such as aberration. An advantage of using \textsc{Geant4} is the possibility to simulate also the particle interaction with materials to obtain the energy loss. Dedicated simulations in the dual-solenoid mode including degraders and blockers can be easily implemented. \par
The nuclear reaction class implemented in the code provides a great advantage to the user for defining the reaction mechanism and to include theoretical cross sections. Most of the possible reaction channels in the RIBRAS facility are included in the code. \par
The feasibility of a Solenoidal Spectrometer in the RIBRAS facility was investigated for the first time. Preliminary geometry models indicate that reactions inside the RIBRAS solenoids are possible. Forthcoming studies both from simulations and experiment are already under development.

\section*{Acknowledgements}
This work was financially  supported by Fundaç\~ao de Amparo a Pesquisa do Estado de S\~ao Paulo (FAPESP) under Grant Nos.~2018/04965-4 and 2016/17612-7. L.E.T. thanks to Conselho Nacional de Desenvolvimento Científico e Tecnológico (CNPq) for the financial support within the Undergrad Research program at IFUSP. G.F.F. thanks to Comissão Nacional de Energia Nuclear (CNEN) for the financial support within the MSc. scholarship program. The authors acknowledges support by the project INCT-FNA (464898/2014-5).

\section*{Declarations}
\textbf{Conflict of Interest:} The authors declare that they have no confict of interest with any organization regarding the material discussed in this manuscript.


\bibliography{references}   % name your BibTeX data base


\end{document}